# Google Test/ Google Mock to Verify Critical Embedded Software

**H.Cheddadi[1,*], S. Motahhir[2], A. El Ghzizal[1]**

[1] Innovative Technologies Laboratory, EST, SMBA University, Fez, Morocco

[2] Engineering, Systems and Applications Laboratory, ENSA, SMBA University, Fez, Morocco

[*]**hafsa.cheddadi@usmba.ac.ma**

**Abstract:**

Critical embedded systems (CES) have become ubiquitous in whether medical, automotive, or industrial. Software failures in such systems are potentially disastrous and could lead to serious consequences not only financially but also life-threatening. However, besides their omnipresence, CES complexity have grown to extreme measures, faced with this increase, manufacturers and suppliers are increasingly interested in effective methods of testing logical correctness and verifying that software parts are error-free.

Software testing is a collection of methods used to detect and correct bugs and faults found in software code. The first stage of software testing is unit testing, a widely used technique where individual units of source code are isolated, often divided up into classes and functions, and tested separately to aid in the verification of the behavior and functionality of the system. In this chapter, an overview of GoogleTest (GTest), a xUnit C++ testing framework will be performed, and a comparison of different available unit test frameworks for use in the C++ language. On account of the complexity provided by critical embedded systems, it is difficult to isolate and test the system's behavior, mocking techniques were considered to enable a real implementation to be replaced with a fake implementation, the replacement allowed to overcome the challenges related to hardware dependencies and external factors. Further, this chapter describes GoogleMock, a part of GoogleTest C++ testing framework that makes the tests run in isolation.

**Keywords:** Critical embedded systems, GoogleTest, GoogleMock, Unit Testing.

**List of Notations and Abbreviations:**

| | |
|---|---|
| **AAA:** Arrange-Act-Assert. | **GMock:** GoogleMock. |
| **C++:** C++ programming language. | **CES:** Critical embedded systems. |
| **GTest:** GoogleTest. | **XML:** Extensible Markup Language. |

## 1. INTRODUCTION:

Critical embedded system applications are designed for systems whose failures or malfunction lead to disastrous consequences for life safety, such as injury or death, financial loss, and environmental damage. Examples of Critical embedded system applications are given in Figure 1. On account of the complexity provided by critical embedded systems, effective methods of detecting and preventing the existence of software failures were considered. It consists of multiple verification levels, stated from the lowest to the highest level, starting from unit testing, integration testing, system testing to acceptance testing. The first level of testing pyramid aims to ensure that the smallest unit does fulfill its functional specification according to its design structure and is independent of other units. A. Bertolino and E. Marchetti [1]define a unit as the smallest component of a software. It could be one to N functions, 1 to N lines of code, or 1 to N objects, the only criteria is the capability of executing independently. The next step in the testing process is integration testing. Regardless of the integration method performed, incremental or otherwise, the connections between entities must be examined to create larger components in order to evaluate their interactions as fully as possible and to ensure that they are working properly with each other and therefore avoid any bugs that may appear as a result of their interaction. System testing involves the entire system embedded in its actual hardware environment and is mainly aimed to uncover requirements errors and verify that the system elements behave properly as it should in real-world use scenarios  and that overall system function and performance has been achieved [1]. Finally, acceptance testing, performed to determine the system's compliance with the requirement specifications and verify if it is has met the required criteria for delivery to end-users. While the detection of bugs in any testing level is desired, they should preferably be detected during unit testing,  Since it forms the base of the testing pyramid, and the costs to fix the bugs are significantly higher if they are found during later testing levels[2].

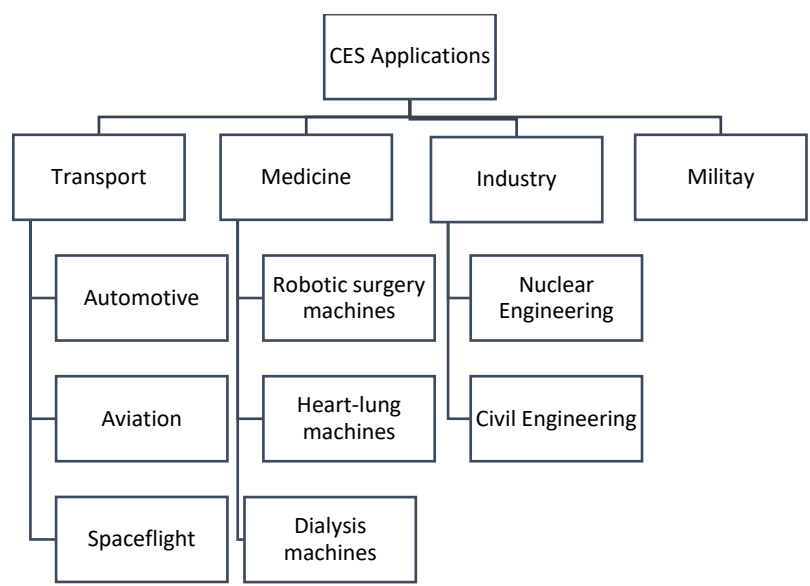

*Figure 1. Critical Embedded Systems Applications.*

The article is organized as follows, the first part is an overview of Google Test, a xUnit C++ testing framework, and regarding the C++ community has a long period of existence and they are different unit testing frameworks, this chapter includes a comparison of selected unit testing frameworks, where the selection criteria was based on popularity and on a range of features. In the second part, a mocking technique (Google Mock), will be considered that helps unit tests run in isolation and overcome challenges related to hardware dependencies and external factors.

## 2. GOOGLE TEST FRAMEWORK

### 2.1 GOOGLE TEST ARCHITECTURE

Google Test is a xUnit C++ testing framework. xUnit is a family of unit-testing frameworks used to write and run repeatable tests for software applications, used for developing and executing unit test cases as well as for regression testing. Amongst the most popular family members of xUnit are JUnit and SUnit which are unit testing frameworks used for testing Java and Smalltalk applications respectively [3]. All the xUnits have the same basic architecture and follow the same design pattern and generally contain five classes: TestCase, TestRunner, TestFixture, TestSuite, and TestResult [4]. Figure 2 shows how they all fit together.

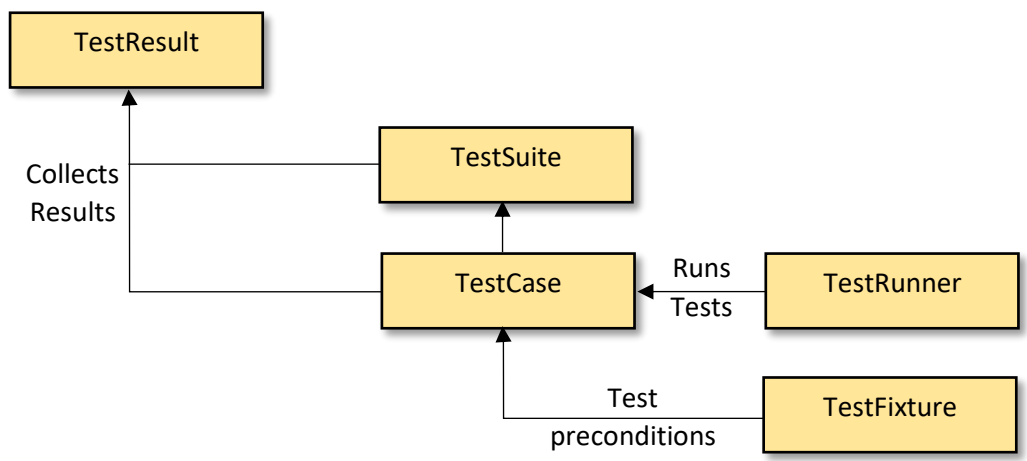

*Figure 2. Main classes of the xUnit test framework architecture.*

In this design, a unit test is structured into a TestCase. The TestSuite acts as a container for TestCase(s). Before a test case is executed, it first gathers the preconditions and data needed. The set of preconditions needed for a test can be shared by multiple test cases using TestFixture. The TestRunner runs tests, saves, and reports the tests results of TestCase assertions in TestResult.

### 2.2 GOOGLE TEST CHARACTERISTICS :

Unit testing has become an important factor in any sort of project success, it enables writing better code faster and without compromising quality. Writing readable, maintainable, sustainable, and robust unit tests is crucial. One of the benefits of using Google Test is that it handles test discovery which means once unit tests are written they are automatically discovered and run by Google Test, it also comes with a build-in runner which enables running the tests as part of an executable created for those tests. Like all other unit testing frameworks, GTest enables running all of the tests or just a subset of them and performs other test run-related operations using the command line arguments.

Google Test provides a wide scope of assertions, different aspects of the results can be checked, from a simple comparison like checking equality or greater to checking whether or not the code under test threw an exception. Google Test also has the ability to run the same test with different parameters, this is an efficient way to create and run multiple tests independently without duplication code. Google Test has the ability to either value parameterized tests, or type parameterized tests. Another Google Test feature, it helps with creating readable and customizable reports at the end of the test run either as a console qoutput or as a JUnit compatible XML file, which then can be consumed by a build server of choice to analyze and show the results.

A unit test should be independent, isolated, and must not affect one another, to ensure that, each test case should be structured according to the Arrange-Act-Assert (AAA) pattern: 'Arrange', initializes objects, preconditions and sets the value of the data that is required for test. 'Act' invokes the method under test with the arranged parameters. Finally, 'Assert' verifies that the expected outcome of the method under test have occurred.

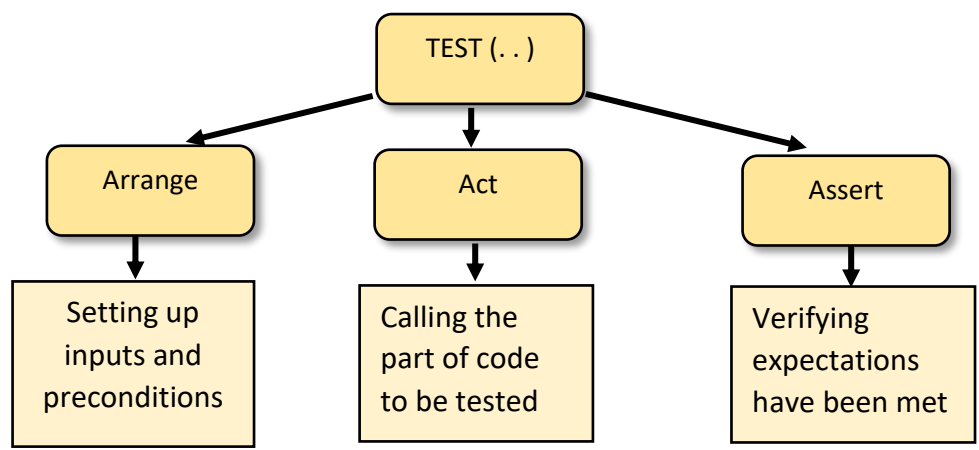

*Figure 3. xUnit testing framework Structure.*

### 2.3 GOOGLE TEST ASSERTION:

Assertions are a Boolean statements that encapsulates a specific logic to check whether a condition is true. An assertion's outcome can be a success, a non-fatal failure, or a fatal failure. Most Assertions are paired with EXPECT_ and ASSERT_ variants. Upon failure, the EXPECT_ macro generates nonfatal failures and allow the current test case to continue its evaluation, while ASSERT_ macros cause fatal failures and termination of test case [5]. These macros are diversified according to the types to be compared: integer, floating, Boolean, strings.

| Fatal assertion | Non fatal assertion | Verifies |
|---|---|---|
| **Binary Comparison** | | |
| ASSERT_EQ(val1, val2) | EXPECT_EQ(val1, val2) | val1 == val2 |
| ASSERT_NE(val1, val2) | EXPECT_NE(val1, val2) | val1 != val2 |
| ASSERT_LT(val1, val2) | EXPECT_LT(val1, val2) | val1 < val2 |
| ASSERT_LE(val1, val2) | EXPECT_LE(val1, val2) | val1 <= val2 |
| ASSERT_GT(val1, val2) | EXPECT_GT(val1, val2) | val1 > val2 |
| ASSERT_GE(val1, val2) | EXPECT_GE(val1, val2) | val1 >= val2 |
| **Boolean Conditions** | | |
| ASSERT_TRUE(condition) | EXPECT_TRUE(condition) | condition is true |
| ASSERT_FALSE(condition) | EXPECT_FALSE(condition) | condition is false |
| **String Comparison** | | |
| ASSERT_STREQ(str1,str2) | EXPECT_STREQ(str1,str2) | Strings have identical content |
| ASSERT_STRNE(str1,str2) | EXPECT_STRNE(str1,str2) | Strings have different content |
| ASSERT_STRCASEEQ(str1,str2) | EXPECT_STRCASEEQ(str1,str2) | Strings have identical content, ignoring case |
| ASSERT_STRCASENE(str1,str2) | EXPECT_STRCASENE(str1,str2) | Strings have different content, ignoring case |

*Table 1. Fatal and non-fatal Assertions.*

With the widespread availability of processors with hardware floating-point units, today's critical systems often need a lot of floating-point computations, so testing or static analysis of programs containing floating-point operators has become a priority [6]. GoogleTest provides assertions for floating-point values that use a default error bound based on Units in the Last Place (ULPs) [5].

| **Floating Points Comparison** | | |
|---|---|---|
| **Fatal assertion** | **Non fatal assertion** | **Verifies** |
| ASSERT_FLOAT_EQ(val1,val2) | EXPECT_ FLOAT_EQ(val1,val2) | The two float values are within 4 ULP's from each other |
| ASSERT_DOUBLE_EQ(val1,val2) | EXPECT_DOUBLE_EQ(val1,val2) | The two double values are within 4 ULP's from each other |
| ASSERT_NEAR(val1, val2,err) | EXPECT_NEAR(val1,str2,err) | The difference between the val1 and val2 doesn't exceed err |

*Table 2. Fatal and non-fatal floating points Assertions.*

Generally, Critical Embedded systems operate in the hard real-time domain, which means that their integrity does not only depend on the functional correctness of the expected behavior, but also on whether these are produced according to a stringent deadline [7].Such systems must be free from unintended and uncontrolled behavior. However, many exceptional conditions can be anticipated when the system is designed, such as bad input for functions, bad memory usage and data errors. Protection against these exception are incorporated into google Test assertions.

| Exception Assertions | | |
|---|---|---|
| **Fatal assertion** | **Non fatal assertion** | **Verifies** |
| ASSERT_THROW(statement, Exception_type) | EXPECT_THROW(statement, Exception_type) | Statement throws an exception of type exception_type. |
| ASSERT_ANY_THROW(statement) | EXPECT_ANY_THROW(statement) | Statement throws an exception of any type. |
| ASSERT_NO_THROW(statement) | EXPECT_NO_THROW(statement) | Statement throws no exception |

*Table 3. Fatal and non-fatal Exception Assertions.*

The assertions above are not the complete list of all the assertions that Google Tests has, there are assertions that check an object is of a specific type, and there are also death tests assertions that verify a statement will cause process to terminate either by exiting with non-zero code or killing the process with a signal. Death tests are considered as sanity checks, and they need to be run in an early stage. As a convention the test case name should end with _DeathTest keyword in order to be run before other tests.

*ASSERT/EXPECT_DEATH (statement, matcher),* Verifies that a program causes the process to terminate and produces an expected error output that correspond 'matcher', That can be a regular expression or a Google Test matcher[8].

*ASSERT/EXPECT_EXIT (statement, predicate, matcher),* Verifies that statement would bring about process termination with an exit status that satisfies predicate, and produces an expected error output that matches matcher. The parameter predicate is a Boolean function that accepts an integer exit status. Google Test provides two predicates to handle abortion on failure cases: ExitedWithCode(exit_code) and KilledBySignal(signal_number) [9].

**2.4 Test Fixtures and Parameterized Tests in GTest:**

Google Test like many other xUnit testing frameworks, supports the notion of fixtures. A test fixture is a fixed state of a set of objects and preconditions used as a baseline for tests in order to refactor the common code out of the tests and reduce code duplication, and in most object-oriented languages, it means grouping the same tests in the same class. Since it is a class, the class fields can be shared between the tests. SetUp() method guaranteed to be run before each and every test case and TearDown() is called afterwards it gives the opportunity to release any resources that may have reserved in the SetUp or during the test.

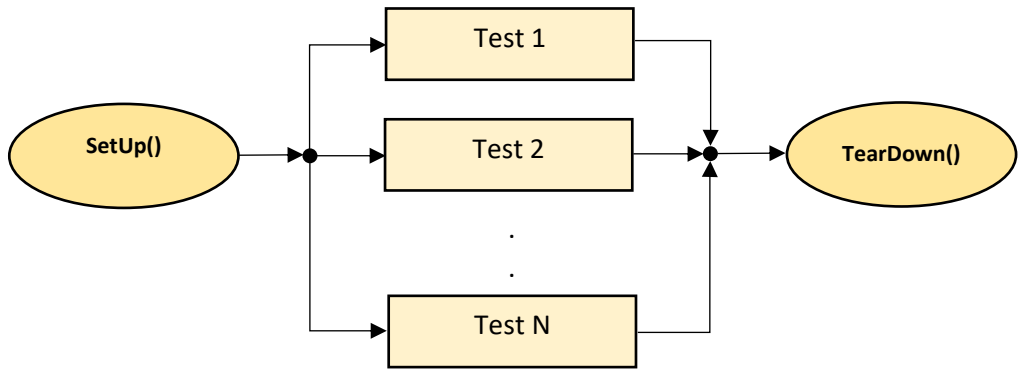

*Figure 4. Test Fixture structure and executing process.*

When using test fixtures, it's important to use *TEST_F()* macro and not *TEST()*, this means for GoogleTest that the class for that test is already defined, and no need to build a new class, this is a class that inherits from *::testing::Test.*

```
Class MyFixture : public ::testing::Test
{
Virtual void SetUp()
   { //Common setup Code }
Virtual void TearDown()
   { // Common TearDown Code }
};
TEST_F(MyFixture, ThisIsTest1)
{
     …
}
.
.
.
TEST_F(MyFixture, ThisIsTestN)
{
     …
}
```

*Figure 5. Test Fixture in Google Test Framework.*

Test fixtures are not the only way to reduce duplication, avoid writing the same test body for different scenarios, is doable using parameterized tests, GTest supports two kind of parameterized tests, Value parameterized tests which can change the value between the test execution using the same code, and type parameterized tests, which change the type of the input for each test. Using parameterized tests, means that I get to write the code once, run it multiple independent times with different values and different parameters reducing code duplication while gaining a lot of power to execute the same scenario with different inputs.

Writing a parameterized test consists of three parts, first of all, create a parameterized test class that inherits from ::testing::TestWithParam<T>. T is a template parameter: the parameter or parameters we want to inject into each iteration. While for a normal unit test TEST() macro and TEST_F() for a fixture, TEST_P() is used for parameterized tests . And lastly INSTANTIATE_TEST_SUITE_P macro is called to instantiate the test suite with different input [10].

| Parameter Generator | Possible Values |
|---|---|
| Range(begin, end [, step]) | {begin, begin+step, begin+step+step, ...}. The end value is excluded. Default step =1. |
| Values(v1, v2, ..., vN) | {v1, v2, ..., vN}. |
| ValuesIn(container) or ValuesIn(begin,end) | values from: a C-style array an STL-style container, an iterator range [begin, end). |
| Bool() | {false, true}. |
| Combine(g1, g2, ..., gN) | combinations of the values generated by the given n generators g1, g2, ..., gN |

*Table 4. Parameter Generator in Google Test.*

```
Class MyFixture : public ::testing::TestWithParams<T>

{

};

TEST_P(MyFixture, ThisIsATest)

{
…
}
INSTANTIATE_TEST_SUITE_P(

            InstantiationName,
            MyFixture,
            ParameterGenerator);
```

*Figure 6. Test Fixture in Google Test Framework.*

## 3. C++ Unit Testing Frameworks:

There are plenty of testing frameworks that exist for such a large language as C++, which has had a long period of existence, all with different features to suit every environment.

In this paper, a brief overview of five different unit testing frameworks was provided, where the selection criteria was based on a range of features and popularity. Table 5 present a summary of different range of features for Doctest, Catch2, CppUnit, Boost Test, and Google Test frameworks.

| Feature | Google Test | Boost Test | CppUnit | Catch2 | Doctest |
|---|---|---|---|---|---|
| **Availability** | Requires building | Header only, static library, or shared library | Requires building | Header Only | Header Only |
| **xUnit** | Yes | Yes | Yes | No | Yes |
| **Assertion** | Yes | Yes | Yes | Yes | Yes |
| **Mocks** | Yes | No | No | No | No |
| **Fixture** | Yes | Yes | Yes | Yes | Yes |
| **Suite** | Yes | Yes | Yes | Yes | Yes |
| **Macros** | Yes | Yes | Yes | Yes | Yes |
| **Exceptions** | Yes | Yes | Yes | Yes | Yes |
| **Death Test** | Yes | Yes | No | No | No |
| **Template** | Yes | Yes | No | Yes | Yes |
| **Modern C++ support** | Yes | Yes | Yes | Yes | Yes |
| **Platform specific Unix Windows OS X** | Yes | Yes | Yes | Yes | Yes |

*Table 5. C++ Unit Testing Frameworks comparison.*

Five unit testing frameworks were identified, most of these frameworks are based on xUnit architecture except Catch2, and all of them support modern C++, handle exceptions, and support assertions and macros. Catch2 is mainly distributed as a header file, but some features may require the inclusion of additional headers. [10]. Google Test and CppUnit require building with the project, Boost Test has three usage variants either header-only, static library, or shared library. Because Doctest and Catch2 are provided as headers, they work across multiple platforms. Only Boost Test and Google Test are the testing frameworks of the five that support death test functionality. CppUnit is the only framework that does not support test template that allows functions and classes to be written to use generic types. Saving time, data and making code easier to read makes the other four testing frameworks stronger candidates. It is also noteworthy to mention, that the testing framework that stands out is Google Test, regarding that it directly supports mocking using Google Mock. The other frameworks lack mocking functionality, moreover, they can be combined with a stand-alone mocking framework such as Google Mock.

## 4. MOCKING FOR UNIT TESTS:

Unit testing critical embedded systems software is not an easy thing to do, very few Functions/methods run in isolation, in many cases, they depend on other functions and methods, or interact with external systems such as external services, databases, third-party software, data centers, which make the testing process more complex [11]. In addition, testers may need to simulate interactions with physical target hardware that may not be available. When testing a unit that has external dependencies, the execution of the test may slow[12], much effort and cost needed to set up for testing [13], and testers should have full control over the external dependencies which is not usually the case, this could be for several reasons including that it has not been developed yet [11]. Moreover, even if the external service was available, after a few tests it will probably blocks testers as they have a limited number of calls they can perform every hour, day, or month. Therefore, including such dependencies in unit testing or regression testing which are usually executed after each committed change will slow down the entire testing process [14].

In unit testing, a software tester should concentrate on the unit under test (e.g. function, component, class) without integrating its dependencies. Including some unstable dependencies may break tests due to bugs in dependencies and make bugs within units harder to identify and fix. Therefore, mock objects were considered to replace real implementation with a fake one that mimics the real dependency and returns the values that the tester tells it to and behaves the way need it to behave for the test to pass. The replacement overcomes challenges linked to hardware/Software dependencies and external factors.

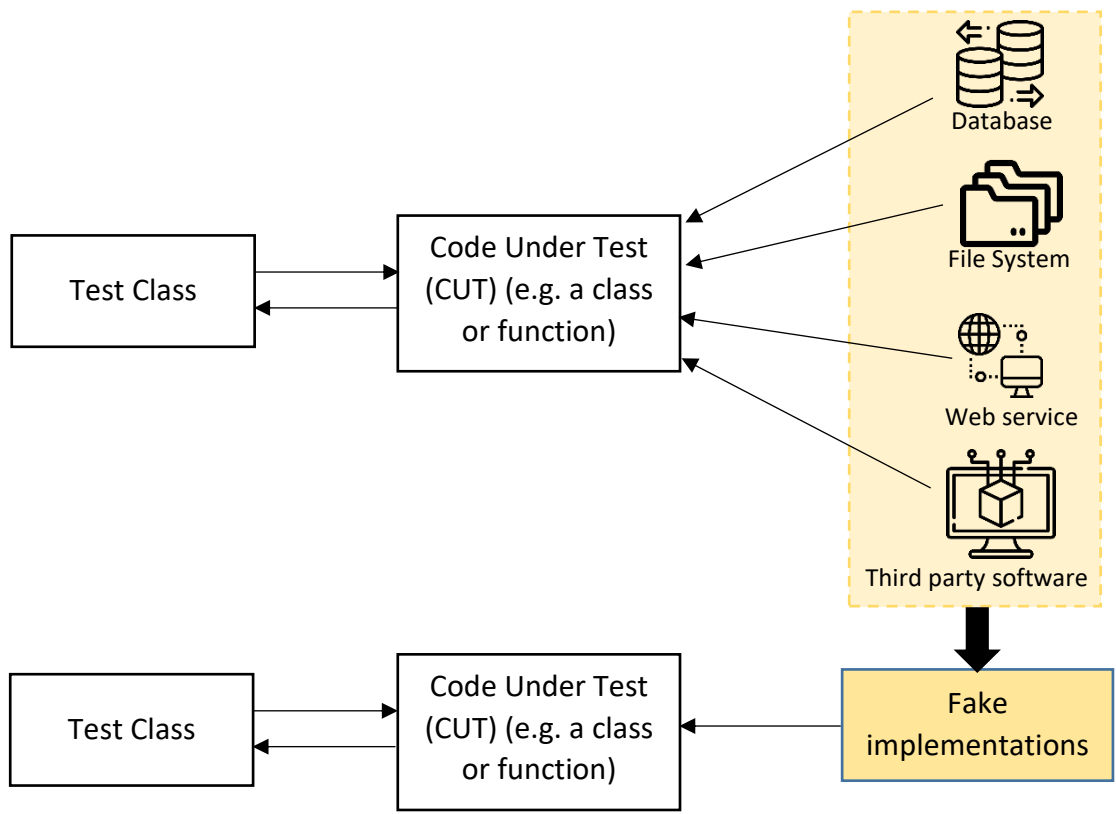

*Figure 7. Isolating Dependencies in Tests Using Fake Implementations.*

hat simulates the behavior of the real objects for testing purposes. There are several types of test doubles, dummy objects, stubs, spies, mocks, and fake objects[15]. One noteworthy point, a lot of people wrap them in the umbrella of all considered as mocks when they are different aspects. According to Langr[16] test doubles are often called mocks, the nuance in terminology is harmless and he summarizes the types of test doubles:

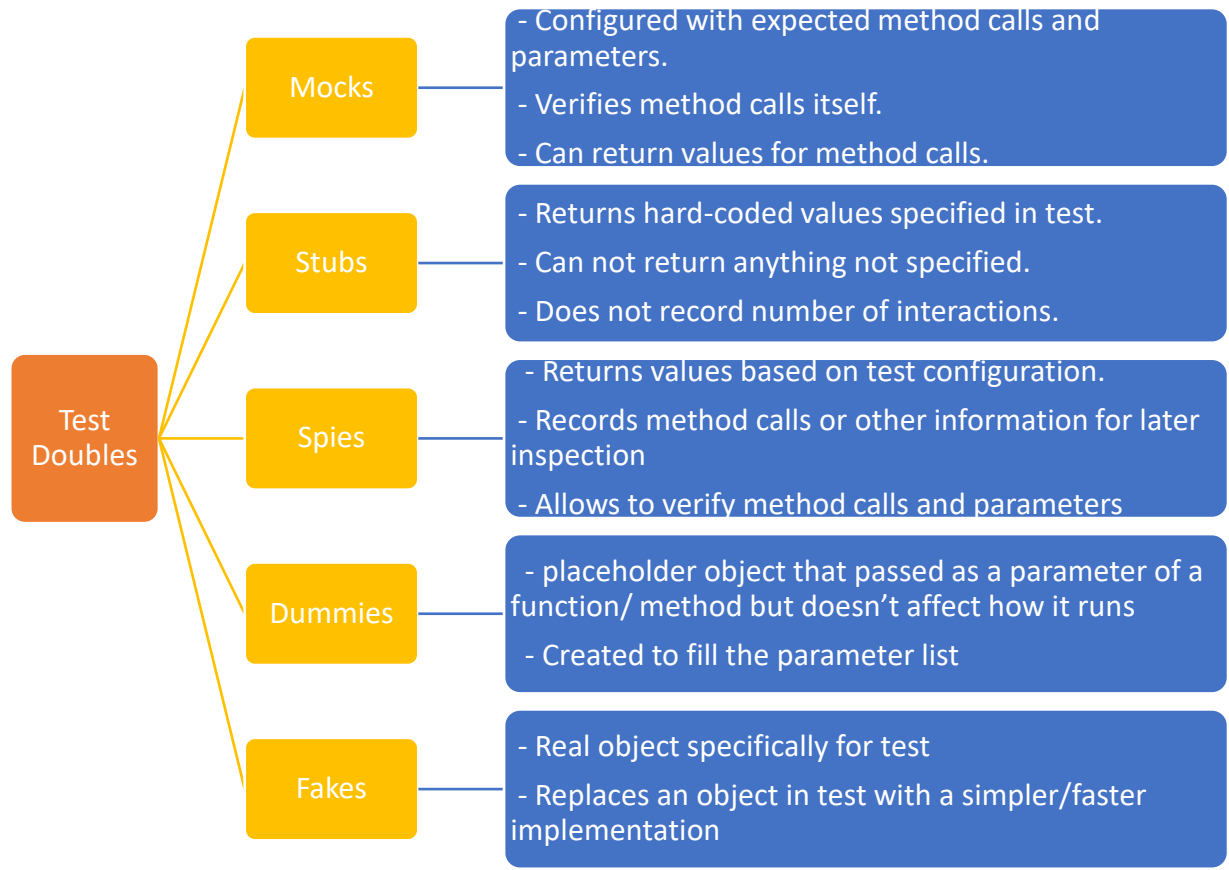

*Figure 8 : Test Double Hierarchy.*

The reason behind using a mocking framework is to reduce the amount of code needed, instead of writing those objects by hand, creating complex logic to satisfy as many tests as possible, it is preferable to use a third-party library to perform most of the heavy lifting for a tester. Mocking is not a C++ specific concept and there are many mocking libraries available for many different languages. However, in this chapter we will be focusing on how to harness the power of mock objects in order to write better unit tests using Google mock library which has been part of Google Test Framework.

Google's C++ Mocking Framework or GMock for short, is an open sourced, free and extendible library for creating fake objects, use them, set behavior on them without the need to change the class with each and every test. This is doable by using the power of inheritance, creating a new derived class, a fake class, and remove the original class functionality by overriding all of its methods and adding empty implementation. Each mocking framework should be able to do at least the following three capabilities: Create fake objects. The second capability is the ability to define a behavior, return a specific value for an exception, this behavior can be defined for each test creating a different scenario each time without affecting other tests. Lastly, each mocking framework can be used to validate if a specific method was called with specific arguments.

### 4.1 CREATING MOCK OBJECTS:

In Google Mock, Creating a mock object requires inheritance from the class we want to mock, the MOCK_METHOD() macro is used, that takes as argument respectively the return value of the method, the name of the method, argument list that the method receives. In case a constant method was mocked, a fourth parameter containing (const) is added [18]. In the backend Google Mock write a new method with an empty Implementation that we can change later on according to the requirements and needs for a test. Obviously for this to work in C++, the following conditions need to be satisfied:

- The Real class must have a virtual destructor as is the case for all classes intend to inherit from.
- All faked methods must be virtual or pure virtual methods so that we'll be able to override them and change their implementation according to the needs of the test.

### 4.2 GOOGLE MOCK: DEFAULT BEHAVIOR:

The whole purpose of creating a fake object is to create an instance that looks like the real class but does absolutely nothing. However, methods have return values, so that fakes methods need to return some value to the test even it is not specified. The default behavior of a method created using GMock are the following: If the method is a void method, obviously no value needs to be returned, if the method has a built-in C++ return type, when it is invoked, the return value will be the equivalent of 0 [19]. In case, GMock cannot return one of the default values, an exception will be thrown telling us that we need to set some behavior on that object either a default one or a specified behavior. There are two ways to change the default return value, either by setting a default value whenever an instance of a type T is returned from a fake object using DefaultValue<T>::Set(value), (reset: DefaultValue<T>::Clear()), if instead it is required to change to a different value for a specific method ON_CALL(mock_object, method).WillByDefault(…) macro is used.

### 4.3 GOOGLE MOCK: SETTING TEST BEHAVIOR:

If the default value returned by google mock doesn't suite the test behavior, an exception will be thrown to set some behavior on that object using the *EXPECT_CALL* macro, where the general syntax is:

```
EXPECT_CALL(mock_object, method(matchers))
       .Times(cardinality)
       .WillOnce(action)
       .WillRepeatedly(action);
```

Using WillOnce(…) means that the behavior will happen only once, if a behavior happens more than once, WillRepeatedly(…) is used. Google Mock support a wide range of actions from simple actions such as returning a value, reference or pointer, throwing an exception, invoking a function to composite actions like using functor or lampda as an action [20].

| Action | Description |
|---|---|
| Return() | Return from a void mock function. |
| Return(value) | Return value. |
| ReturnArg<N>() | Return the N-th (0-based) argument. |
| ReturnNew<T>(a1, ..., ak) | Return new T(a1, ..., ak); a different object is created each time. |
| ReturnNull() | Return a null pointer. |
| ReturnPointee(ptr) | Return the value pointed to by ptr. |
| ReturnRef(variable) | Return a reference to variable. |
| ReturnRefOfCopy(value) | Return a reference to a copy of value; the copy lives as long as the action. |
| f | Invoke f with the arguments passed to the mock function, where f is a callable. |
| Invoke(f) | Invoke f with the arguments passed to the mock function, where f can be a global/static function or a functor. |
| Invoke(object_pointer, &class::method) | Invoke the method on the object with the arguments passed to the mock function. |
| InvokeWithoutArgs(f) | Invoke f, which can be a global/static function or a functor. f must take no arguments. |
| InvokeWithoutArgs(object_pointer, &class::method) | Invoke the method on the object, which takes no arguments. |
| InvokeArgument<N>(arg1, arg2, ..., argk) | Invoke the mock function's N-th (0-based) argument, which must be a function or a functor, with the k arguments. |
| Throw(exception) | Throws the given exception |

*Table 6 : Google Mock Actions[20].*

### 4.4 GOOGLE MOCK: VERIFYING BEHAVIOR:

Beside the fact that EXPECT_CALL enables setting behavior on fake object and defining what should happen when a method is called either invoke a different method, return a value, or throw an exception, it also tells google mock to verify that the method we expecting to be called did in fact occurred during a unit test run, otherwise, the test will fail. There are several ways to affect how our EXPECT_CALL behaves.

```
EXPECT_CALL(mock_object, method(matchers))
        .With( multi arguments matchers)
        .Times(cardinality)
        .InSequence(S1...Sn)
        .After( other calls expectation)
        .WillOnce(action)
        .WillRepeatedly(action)
        .RetiresOnSaturation();
```

*EXPECT_CALL* has two parameters the mock object and the method we're expecting to be called, as parameters to that method, explicit arguments, underscore, or special objects (matchers) can be passed. We can specify additional multi-argument matchers using *'With'*. '*Times'* can be added to explicitly specify how many times we expect a method to be called. Since google mock expectation are stack based: the last expectation is the most relevant one, *'InSequence'* ensures a specific sequence order to satisfy a test and that the test should fail if the method will be called out of the defined order. Similarly using *'After'* tells google mocks that this call should be done after another calls. To control expectations lifecycle *'RetiresOnSaturation'* is used that tells Google Mock to retire the call as it gets saturated. The full calls stack is rarely used and testers usually define only the parts needed for a specific unit test.

Using mocking frameworks in unit testing leads to classify unit tests in two types: In one hand, state Base Testing "classic testing" follows the structure *Arrange-Act-Assert (AAA):* scenario set up, method executed, expectation checked against returned value or object state using assertions. On the other hand, interaction testing "Mockist" , where we have the option to either test the object state just like classic unit testing or verifying if methods were either called or not. Therefore, we perform interaction testing using Google Mock in the following cases: First, test result is not accessible. Second, test result is external to the system under test. The last reason is testing the behavior is part of a business requirement.

## 5. CONCLUSION

In summary, this paper aimed to develop a guideline encircle a set of C++ unit testing tools. The identified frameworks were Google Test, Boost Test, CppUnit, Catch2 and Doctest, though Google Test is a stronger candidates for testing critical embedded software since it supports a wide large set of assertions, different aspects of the results can be checked, and it directly support mocking by replacing the real dependencies with simple implementation using Google Mock.

## 6. FIGURES AND TABLES CAPTION LIST

Figure 1. Critical Embedded Systems Applications.

Figure 2. Main classes of the xUnit test framework architecture.